\documentclass[a4paper, 11pt]{elsarticle}

\usepackage[margin = 2.5cm]{geometry}
\usepackage{amssymb}
\usepackage{listings}
\usepackage{float}
\newfloat{Listing}{tbh}{lot}

\usepackage{tikz}
\usetikzlibrary{shapes.geometric, arrows.meta, matrix, positioning, fit}

\usepackage[colorlinks=true, linkcolor=black, citecolor=blue, urlcolor=blue, unicode=true]{hyperref}

\begin{document}
\renewcommand{\labelenumii}{\arabic{enumi}.\arabic{enumii}}

\begin{frontmatter}

\title{Magpie: A Graphical Magnon Software}

\author{Tobias Weber}
\ead{tweber@ill.fr}
\address{Institut Laue-Langevin (ILL), 71 avenue des Martyrs, 38000 Grenoble, France}

\begin{abstract}
\textit{Magpie} is a fully graphical linear spin-wave software which has its main focus on
3-d visualisation of magnetic models and magnon dispersion relations.
The software is distributed under an open-source license and is written in high-performance, portable C++.
Its modular architecture allows for a rapid development of new calculation modules
and permits the inclusion into external applications.
Included code generators ensure seamless interoperability with other linear spin-wave libraries.
\end{abstract}


\end{frontmatter}


\section*{Required Metadata}

\section*{Current code version}

\begin{table}[!h]
\begin{tabular}{|l|p{6.5cm}|p{6.5cm}|}
\hline
\textbf{Nr.} & \textbf{Code metadata description} & \\
\hline
C1           & Current code version		& Version 1.2 \\
\hline
C2           & Permanent GitHub link to code/repository used for this code version
	& \url{https://github.com/ILLGrenoble/magpie} \\
\hline
C3           & Legal code license		& GNU General Public License version 3 \\
\hline
C4           & Code versioning system used	& git \\
\hline
C5           & Software code languages, tools, and services used
	& C++20,
	\href{https://www.python.org}{Python},
	\href{https://cmake.org}{CMake},
	\href{https://www.swig.org}{SWIG},
	\href{https://www.boost.org}{Boost},
	\href{https://www.netlib.org/lapack}{Lapack(e)},
	\href{https://github.com/project-gemmi/gemmi}{Gemmi},
	\href{http://www.qhull.org}{QHull},
	\href{https://root.cern}{Minuit2},
	\href{https://www.qt.io}{Qt},
	\href{https://www.qcustomplot.com}{QCustomPlot} \\
\hline
C6           & Compilation requirements, operating environments \& dependencies
	& Linux, MacOS, and Windows (cross-compilation via \href{https://www.mingw-w64.org}{MinGW}) \\
\hline
C7           & Link to developer documentation/manual
	& \url{https://github.com/ILLGrenoble/magpie/wiki} \\
\hline
C8           & Support email for questions	& \url{tweber@ill.fr} \\
\hline
\end{tabular}
\caption{Code metadata.}
\label{tab:metadata}
\end{table}

\section{Motivation and significance}
Many software systems exist for linear spin-wave calculations, most prominently among them
\textit{SpinW} \cite{Toth2015}, \textit{Sunny} \cite{Sunny}, and \textit{McPhase} \cite{Rotter2004}.
Even the software presented here has a published predecessor, 
namely the magnon plugin of the neutron data analysis software \textit{Takin} \cite{Takin2023}. 
At the time it was developed, \textit{Sunny} had not yet been available, 
interfacing with the proprietary language that \textit{SpinW} was developed in was unfeasible, 
and \textit{McPhase} was generally too complex to use.
Thus, there was a strong motivation at that time to develop an in-house,
minimalistic implementation of linear spin-wave theory featuring a rudimentary graphical user interface (GUI).

After three years of subsequent development, \textit{Magpie} is now the evolution of this originally 
very compact magnon module into its own fully featured application.
Its main motivations are ease of use and performance.
Ease of use is achieved by providing all functionality for creating and visualising magnetic models
and magnon dispersion relations via a graphical user interface (GUI), both in two and in three dimensions.
This feature distinguishes the programme from other linear spin-wave calculators which rely entirely on scripting.
Performance is achieved by developing the programme as a native application entirely using C++.

\section{Software description}
\textit{Magpie} is a fully graphical magnon calculation software solving Hamiltonians of the form
\begin{equation}
	\mathcal{H} =
		\sum_{i \neq j}  \mathbf{S}_i J_{ij} \mathbf{S}_j
		+ \sum_{i \neq j} \mathbf{D}_{ij} \cdot \mathbf{S}_i \times \mathbf{S}_j 
		+ \sum_i \mathbf{S}_i A_i \mathbf{S}_i
		+ \mu_B \mathbf{H} \sum_i g_i \mathbf{S}_i,
\end{equation}
where $J_{ij}$ is the symmetric Heisenberg exchange interaction between two spins $\mathbf{S}_i$ and $\mathbf{S}_j$.
The $J_{ij}$ can be either scalars or matrices.
$\mathbf{D}_{ij}$ denotes the antisymmetric Dzyaloshinskii-Moryia interaction (DMI) vector.
$A_i$ is the single-ion anisotropy (SIA) matrix coupling spin $\mathbf{S}_i$ with itself.
In practice, the $J_{ij}$ are always treated as matrices, where the exchange interactions are given as diagonal elements 
and the DMI as off-diagonal skew-symmetric elements.
SIA is treated in the exact same way, only with the spin $\mathbf{S}_i$ coupling to itself.
$\mathbf{H}$ is the external magnetic field, with Bohr magneton $\mu_B$ and electron g-factor $g_i$.
Determining the eigensystem of the Hamiltonian is done using the well-known framework of linear spin-wave theory \cite{Toth2015, McClarty2022},
which is shortly summarised in the following paragraph.
In linear spin-wave theory, the general case of non-collinear classical spin directions is reduced to the ferromagnetic case of collinear spins
by local transformations at each magnetic site.
In these local frames, the spin operators $\mathbf{S}_i$ are approximated in terms of boson creation and annihilation operators, 
$b^\dagger$ and $b^-$, via a Holstein-Primakoff transformation.
Ensuing roots in the Hamiltonian are expanded up to quadratic order in $S$, the magnitude of the classical spin vectors, 
with the linear order vanishing in the collinear ground state \cite{McClarty2022}.
Fourier transforming the boson operators, 
$b_i = 1/\sqrt{N} \cdot \sum_{\mathbf{q}} \exp \left( \imath \mathbf{q} \cdot \mathbf{r}_i \right) b_{\mathbf{q}}$, 
introduces the momentum transfer $\mathbf{q}$, where the $\mathbf{r}_i$ are the positions of the magnetic sites.
The Hamiltonian is diagonalised by a Bogoliubov transformation of the boson ladder operators. 
The transformation is needed because the local rotations introduced for non-collinear systems would otherwise cause the boson operators 
to not satisfy the canonical commutation relations.
More information as well as a constructive step-by-step guide through the theoretical linear spin-wave model can be found in Ref. \cite{Toth2015}, 
which is also the main basis of the implementation of the present calculation kernel.

In the software, the magnetic system's spin structure $\mathbf{S}_i$ can be built and visualised in three dimensions using \textit{OpenGL}.
The software allows the user to position magnetic sites and the interactions between them.
At the same time, the influence of the structure and the coupling constants on the dynamics is visualised live and interactively.
The magnon dispersion is shown both in a two-dimensional and in a three-dimensional cut 
of the dynamical structure factor $S\left(\mathbf{Q}, E \right)$.
To ease the navigation in reciprocal space, the first nuclear Brillouin zone and nuclear Bragg peaks are rendered 
and the current dispersion path displayed.

\subsection{Software architecture}
\begin{figure}[htb]
\begin{centering}
	\begin{tikzpicture}[node distance = 1.2cm]
		\tikzstyle{container} = [ matrix of nodes, nodes = redbox, column sep = 0.1cm, row sep = 0.1cm ]
		\tikzstyle{normal} = [ rectangle, rounded corners,
			minimum width = 2.5cm, minimum height = 0.75cm,
			draw = black, fill = white, text centered, ]
		\tikzstyle{nofill} = [ rectangle, rounded corners,
			minimum width = 2.5cm, minimum height = 0.75cm,
			draw = black, fill = none, text centered, ]
		\tikzstyle{redbox} = [ rectangle, rounded corners,
			minimum width = 2.5cm, minimum height = 0.75cm,
			draw = black, fill = red!50, text centered, ]
		\tikzstyle{graybox} = [ rectangle, rounded corners,
			minimum width = 2.5cm, minimum height = 0.75cm,
			draw = black, fill = gray!50, text centered, ]
		\tikzstyle{nobox} = [ draw = none, color = black, fill = none, minimum height = 0cm ]

		\matrix(magdyn)[container]{
			|[nobox, yshift = 0.25cm]| \textit{Magpie} core \\
			Spin-wave library & Symmetry library \\
			Form factor parser & Code generators \\
		};
		\node(magdyn2)[nofill, fit = (magdyn)] { };

		\matrix(libs)[container, xshift = -6.25cm]{
			|[nobox, yshift = 0.25cm]| Helpers \\
			3-d library \\
			Brillouin zone library \\
		};
		\node(libs2)[nofill, fit = (libs)] { };

		\node(magpie)[normal, below of = libs, xshift = -1.1cm, yshift = -1.5cm]{\textit{Magpie} GUI App};
		\node(magpy)[normal, right of = magpie, xshift = 2.3cm, yshift = 0cm]{\textit{Python} interface};
		\node(clib)[normal, right of = magpy, xshift = 2cm, yshift = 0cm]{\textit{C} interface};
		\node(takin)[graybox, right of = clib, xshift = 2cm]{Plugin for \textit{Takin}};

		\draw[thick, -Triangle] (magpy) to (magdyn);
		\draw[thick, -Triangle] (magpie) to (magdyn);
		\draw[thick, -Triangle] (magpie) to (libs);
		\draw[thick, -Triangle] (clib) to (magdyn);
		\draw[thick, -Triangle] (takin) to (magdyn);
	\end{tikzpicture}

	\caption{Software architecture comprising the \textit{Magpie} core, helper libraries, as well as the main
	GUI application, the \textit{Python}, and the \textit{C} scripting interfaces.
	As example for an external application, the magnon plugin module for \textit{Takin} \cite{Takin2023}
	is also shown.}
	\label{fig:architecture}
\end{centering}
\end{figure}
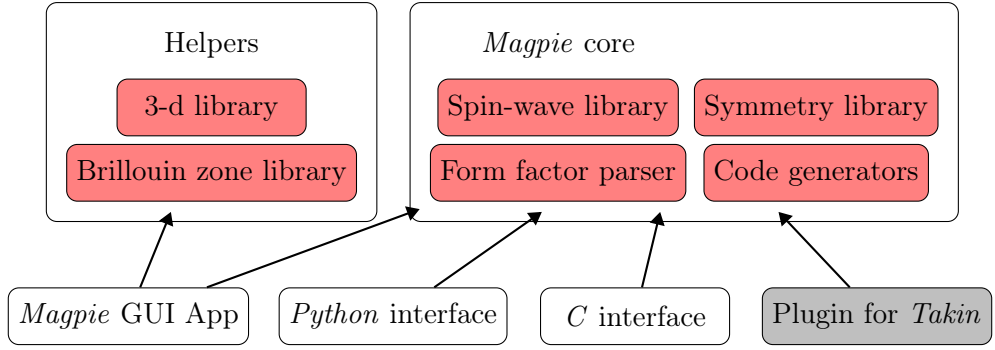

The \textit{Magpie} software is designed in a strongly modular fashion, see Fig. \ref{fig:architecture}.
It comprises a core library, a graphical user interface (GUI), see Fig. \ref{fig:screenshot},
a C++ and a C application programming interface (API), as well as a \textit{Python} scripting interface (``\textit{Magpy}'').
\textit{Magpie} grew out of the -- at that time -- very minimalistic magnon calculation module that was originally developed
for the \textit{Takin} neutron scattering software \cite{Takin2023}.

The core module is realised as a C++20 template library and builds on a further developed \textit{tlibs2}
library that was introduced with the \textit{Takin} \cite{Takin2021, Takin2023} and the \textit{TAS-Paths} \cite{TasPaths2023}
software suites.
Low-level linear algebra calculations are performed using Netlib's reference implementation of \textit{Lapack} and \textit{BLAS} \cite{Lapack}.
We found that in our case these reference implementations often out-perform other optimised implementations.
The reason is that \textit{Magpie} manages its own threading and as such prefers single-threaded low-level libraries.
Threading in \textit{Magpie} is done by queuing calculations into a pool and is implemented using \textit{Boost.Asio} \cite{BoostAsio}.

\textit{Magpie's} core module can be used directly as an easy-to-include header-only C++ library.
Furthermore, a pure-C wrapper library is available, as well as a \textit{Python} scripting interface.
The latter is realised using the \textit{SWIG} interface generator \cite{SWIG}.
While on \textit{Magpie's} interface side everything is exposed,
at the moment certain functionalities are limited because
of SWIG's limited support for certain C++20 features like template concepts.
To still ensure access to these functionalities the interface module provides helper functions.

The main feature of the \textit{Magpie} suite is its graphical user interface (GUI), which is realised
using the \textit{Qt} library \cite{Qt}, the \textit{QCustomPlot} \cite{QCustomPlot} plotter for two-dimensional displays,
as well as a custom-written \textit{OpenGL} plotter for three-dimensional visualisations.
The \textit{OpenGL} plotter is a further development of the \textit{OpenGL} renderer used
mainly in the \textit{TAS-Paths} software \cite{TasPaths2023}.
Furthermore, the Brillouin zone module originating from the \textit{Takin} software \cite{Takin2023} has been further developed
for \textit{Magpie}. 
Internally, Brillouin zone calculation relies on the \textit{QHull} \cite{Qhull} library.

Next, the functionalities of the principal modules are explained.

\subsection{Software functionalities}
\textit{Magpie's} core module comprises the functions needed to build magnetic models and calculate their dynamics
using linear spin-wave theory \cite{Toth2015}.
The basic workflow is as follows:
First, magnetic site positions and the possible exchange paths between them are created up to a given order
using the symmetry operations of the crystal's space groups, which are obtained from the \textit{Gemmi} library \cite{Wojdyr2022}.
Second, system's ground state spin configuration is determined by minimising the system's energy
using the \textit{Root/Minuit2} library \cite{Root}.
Finally, magnetic dynamics comprising the dispersion relation and the spin-spin correlation function
can be calculated for single-crystals and for powders.
In the latter case, the spectrum is averaged from the results obtained by calculating the dispersion on Monte-Carlo positions
created on a sphere in reciprocal space whose radius corresponds to the momentum transfer.

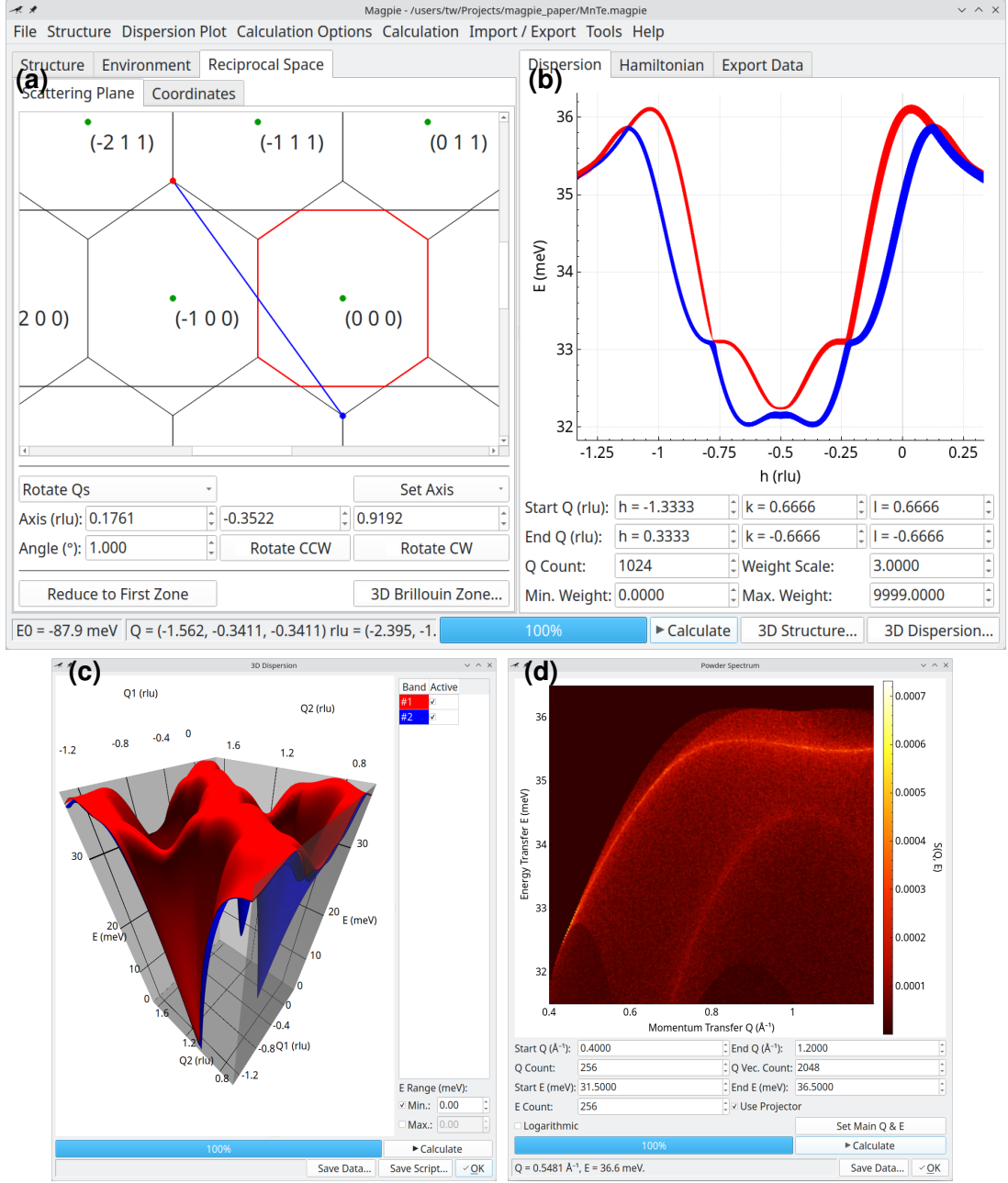
\begin{figure}[htb!]
\begin{centering}
\begin{tikzpicture}
	\draw (0, 0) node [ inner sep = 0 ] { \includegraphics[width=0.9\textwidth]{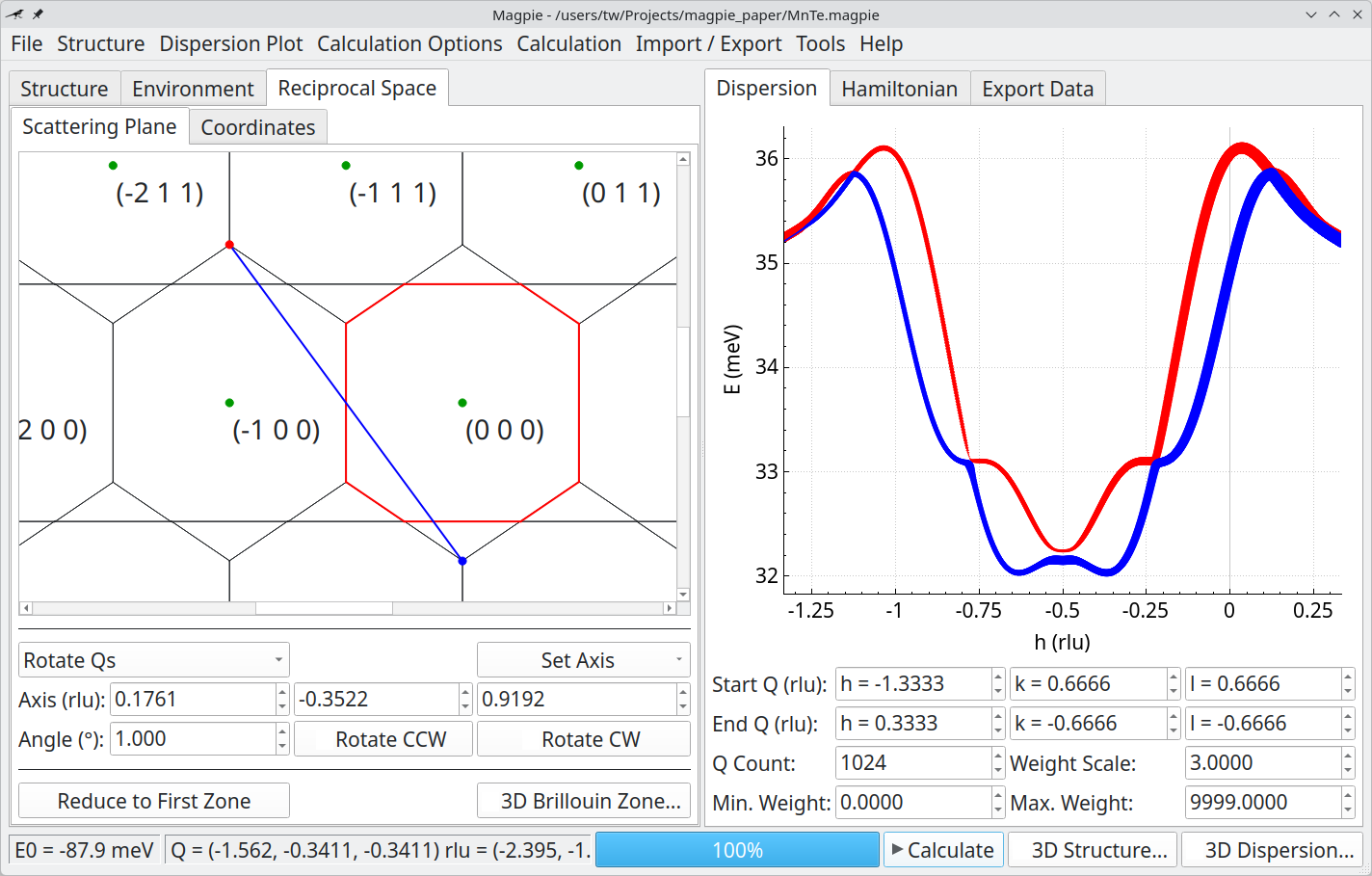} };
	\draw (-6.9, 3.5) node { \bf \fontfamily{phv} \selectfont (a) };
	\draw (0.5, 3.5) node { \bf \fontfamily{phv} \selectfont (b) };
\end{tikzpicture}
\begin{tikzpicture}
	\draw (0, 0) node [ inner sep = 0 ] { \includegraphics[width=0.4\textwidth]{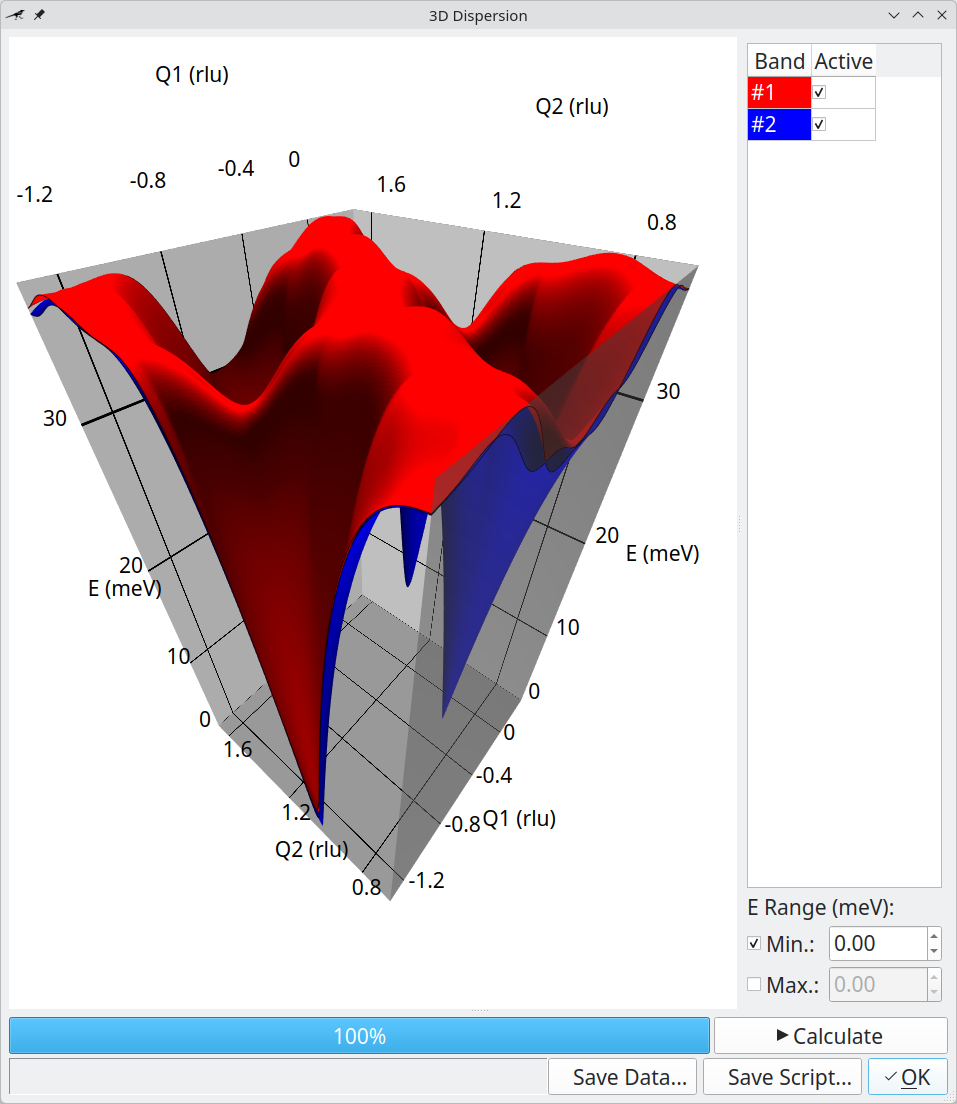} };
	\draw (-2.8, 3.55) node { \bf \fontfamily{phv} \selectfont (c) };
\end{tikzpicture}
\begin{tikzpicture}
	\draw (0, 0) node [ inner sep = 0 ] { \includegraphics[width=0.4\textwidth]{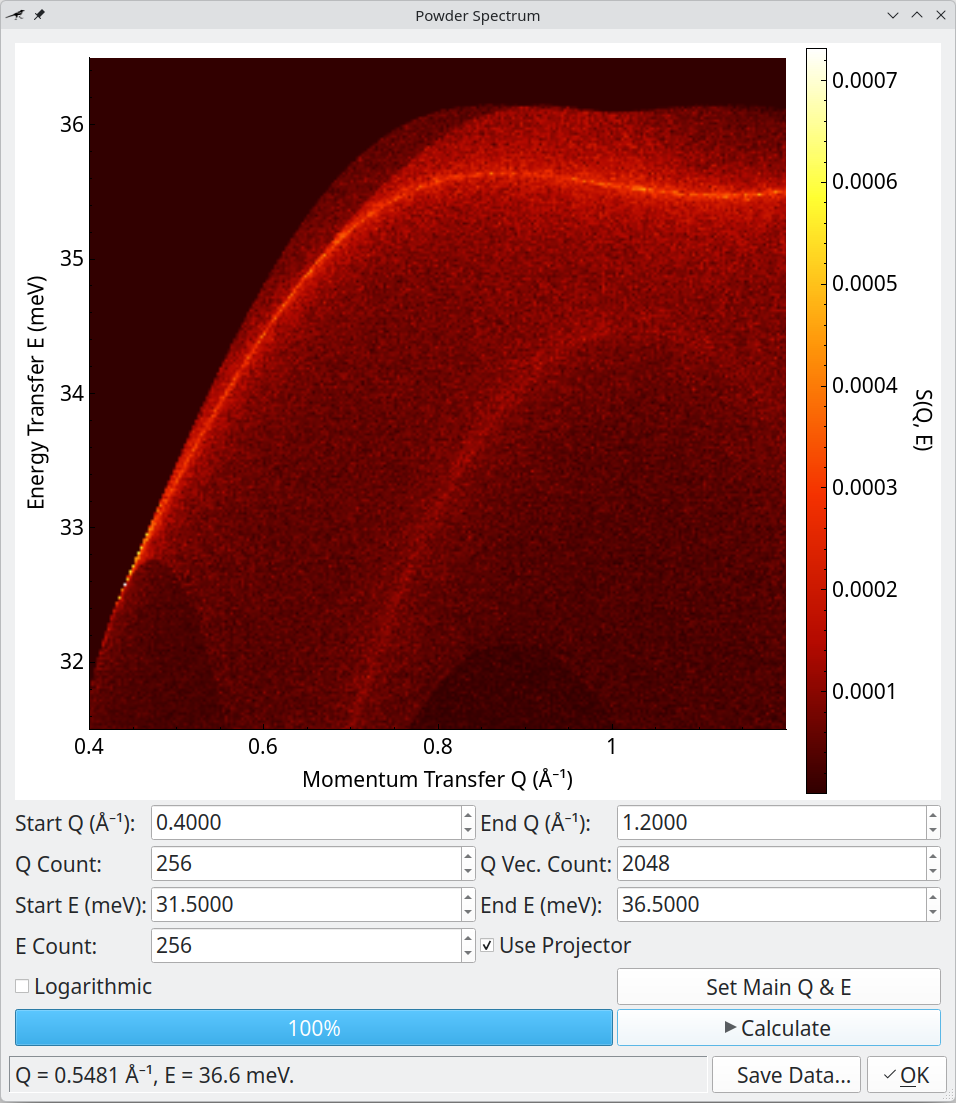} };
	\draw (-2.8, 3.55) node { \bf \fontfamily{phv} \selectfont (d) };
\end{tikzpicture}
	\caption{Typical \textit{Magpie} session showing a selection of the software's calculation modules.
		The magnetic structure in this example is altermagnetic MnTe \cite{Liu2024, mnte}.
		Panel (a) shows a 2-d cut of the reciprocal space using an $\left(hkk\right)$ scattering plane.
		In this plane, the accessible Bragg peaks are drawn as green dots
		and the crystal's nuclear Brillouin zone as black and red lines.
		The currently selected scan path is depicted as a blue line.
		Panel (b) and (c) show the 2-d and 3-d renditions of the magnon dispersion relation, respectively,
		corresponding to the scan path of panel (a).
		The display colour-codes the magnon bands, here in red and blue.
		The scan path is also indicated as grey plane in panel (c).
		Panel (d) shows the corresponding calculated powder spectrum for the same crystal.}
	\label{fig:screenshot}
\end{centering}
\end{figure}

All of the functionalities of the core module are accessed graphically via the GUI module as shown in Fig. \ref{fig:screenshot}.
The GUI lets the user build magnetic models and choose paths in reciprocal space
for which the dispersion relation is calculated via a point-and-click interface.
The scan paths in reciprocal space are rendered together with a two- and three-dimensional representations
of the crystal's Brillouin zone (Fig. \ref{fig:screenshot} (a)),
which is calculated via a Voronoi diagram of the available nuclear Bragg reflections.

Results are visualised in both two- and three dimensional plots, 
as depicted in Fig. \ref{fig:screenshot} (b) and (c) for a single-crystal, 
and Fig. \ref{fig:screenshot} (d) for a powder.
All three-dimensional visualisations are interactive in that they let the user click on an element,
for example an exchange path in the structure viewer or a magnon band in the dispersion plot,
and get information on the object and the options to manipulate its properties.
The user can, for example, graphically remove an exchange path or hide a magnon band, respectively.
Such kind of interactivity aids in the understanding of complicated dispersions,
as the user can selectively hide uninteresting parts of the dispersion directly in the three-dimensional view.

\textit{Magpie} features code generators that allow to export the magnetic model towards other linear spin-wave toolkits,
among these \textit{SpinW} \cite{Toth2015} and \textit{Sunny} \cite{Sunny}.
Furthermore, it is possible to generate a self-contained \textit{Python} code,
although this last feature is limited to non-commensurate magnetic models for the moment.
This kind of interoperability with other toolkits allows for a high flexibility when analysing magnetic systems. 
It furthermore permits an easy verification of the consistency of results.

Results can be directly exported to \textit{Python} scripts using \textit{Matplotlib} \cite{Matplotlib}.
This allows for the easy creation of publication-ready plots of the entire band structure in either two or three dimensions.
In the latter case, an export to the \textit{Mayavi} \cite{Mayavi} software is also possible.

The modular nature of the software allows for an easy inclusion of further calculation modules.
For example, a modules exist for the calculation and visualisation of magnon group velocities as well as 
a currently still preliminary module for the analysis of the topological properties of the magnon band structure,
the latter using the formalism given by McClarty \cite{McClarty2022}.

\subsection{Sample code snippet analysis}
Although \textit{Magpie} is mainly intended to be used via its GUI, it is possible to script the workflow
using \textit{Python}. In Listing \ref{lst:scripting} we show a simple script calculating the dispersion of altermagnetic MnTe.

\begin{Listing}[htb]
	\begin{lstlisting}[basicstyle=\tiny, tabsize=4, language=Python, numbers=left]
import magpy, numpy
mag = magpy.MagDyn()  # create a new magpie object

spacegroup = "P 63/m m c"
mag.SetCrystalLattice(4.1, 4.1, 6.6,
	90./180.*numpy.pi, 90./180.*numpy.pi, 120./180.*numpy.pi)

# add coupling variables (values from https://arxiv.org/abs/2408.16490)
magpy.add_variable(mag, "J_afm1", 3.99)
magpy.add_variable(mag, "J_afm2", 0.47)
magpy.add_variable(mag, "J_fm1", -0.12)
magpy.add_variable(mag, "J_am1",  0.068)
magpy.add_variable(mag, "J_am2", -0.022)

# set up the sample environment
magpy.set_temperature(mag, 50.)

# create magnetic sites
#                   name    x    y    z     sx   sy   sz    S
magpy.add_site(mag, "Mn",   0.,  0.,  0.,   0., +1.,  0.,   5./2.)
magpy.symmetrise_sites(mag, spacegroup)
magpy.set_spin(mag, "1", 0., -1., 0.)

# create couplings between sites
#                       max. distance  supercell  count
magpy.generate_couplings(mag,     10.,         5,   200, spacegroup)

# assign coupling constants by symmetry index, sorted by length
magpy.assign_couplings_by_symmetry(mag,  1, "J_afm1")
magpy.assign_couplings_by_symmetry(mag,  2, "J_fm1")
magpy.assign_couplings_by_symmetry(mag,  3, "J_afm2")
magpy.assign_couplings_by_symmetry(mag, 10, "J_am1")
magpy.assign_couplings_by_symmetry(mag, 11, "J_am2")
mag.RemoveUnusedExchangeTerms()

magpy.calc(mag)  # calculate sites and couplings

# calculate the dispersion and write it to a file
mag.SaveDispersion("dispersion.dat",
	1./3., 0., 0.,  # (hkl) to calculate from
	1./3., 0., 2.,  # (hkl) to calculate to
	128)            # number of Q points
	\end{lstlisting}
	\caption{\textit{Python} code for altermagnetic MnTe,
		reproducing the model and using the coupling constants from \cite{Liu2024}.}
	\label{lst:scripting}
\end{Listing}

Lines 4 to 6 of the script set up the crystal lattice.
In lines 8 to 13, some variables are defined, which will be used as coupling constants.
Symmetry-equivalent magnetic sites are created by the crystal's space group.
For that, one position is explicitly specified in line 20, the rest is generated in line 21.
All possible magnetic couplings up to a given maximum length are generated in line 26.
The previously defined coupling constants are assigned to five classes of symmetry-equivalent couplings in lines 28 to 33.
In the final step, a dispersion branch is calculated in lines 38 to 42.

\section{Illustrative example}
\label{sec:example}
Several examples are given in the software's documentation in the form of tutorials\footnote{\url{https://github.com/ILLGrenoble/magpie/wiki}}.
Here, we show how to create the same structure and dynamics as in the script given in Listing \ref{lst:scripting}, but using the GUI.
This example is also available as a video tutorial\footnote{\url{https://github.com/ILLGrenoble/magpie/wiki/Video-Tutorials}}.

The system for which we illustrate the functionalities of \textit{Magpie} is altermagnetic $\alpha$-MnTe, 
We demonstrate how to recreate the theoretical model that was originally conceived by Liu \textit{et al.} and fitted experimentally \cite{Liu2024}.

First, the crystal's lattice constants and angles, as well as the space group, 
have to be entered, for MnTe these are $a = b = 4.1\ \mathring{\mathrm{A}}$,
$c = 6.6\ \mathring{\mathrm{A}}$ and $\alpha = \beta = 90^{\circ}$, $\gamma = 120^{\circ}$, 
and space group \#194, respectively \cite{Liu2024}.
This serves to calculate the metric tensor of the crystal coordinate system and calculate the coupling lengths 
of the magnetic exchanges in a later step.
It furthermore serves to calculate the orthogonal projection operator for the transverse component of 
the dynamical structure factor seen by neutron scattering.

Second, the magnetic sites are specified. 
Here, we have a Mn position at $\left[000\right]$ with spin direction along $\left[010\right]$ 
and a total spin magnitude of 5/2 \cite{Liu2024}.
Symmetry-equivalent sites are generated by the space group.
This corresponds to one additional position, for which we manually flip the spin direction, 
which is not considered in the nuclear space groups.

Finally, the magnetic couplings are created.
We select a maximum distance so that the neighbours of up to 11th order are included. 
These long-distance interactions are what distinguishes an altermagnet from a conventional antiferromagnet \cite{Liu2024}.
We assign the exchange constants by symmetry index, i.e. 
$1 \rightarrow 3.99\ \mathrm{meV}$, 
$2 \rightarrow -0.12\ \mathrm{meV}$,
$3 \rightarrow 0.48\ \mathrm{meV}$,
$10 \rightarrow 0.068\ \mathrm{meV}$,
$11 \rightarrow 0.022\ \mathrm{meV}$,
where the left hand side gives the symmetry index sorted by length, 
and the right-hand side specifies the experimentally-determined \cite{Liu2024} exchange constants.
These constants can either be input numerically, as shown here, or registered as a variable.
The latter would be needed in case the model is used in a fitting procedure, 
for example using the triple-axis resolution software \textit{Takin} \cite{Takin2023}.

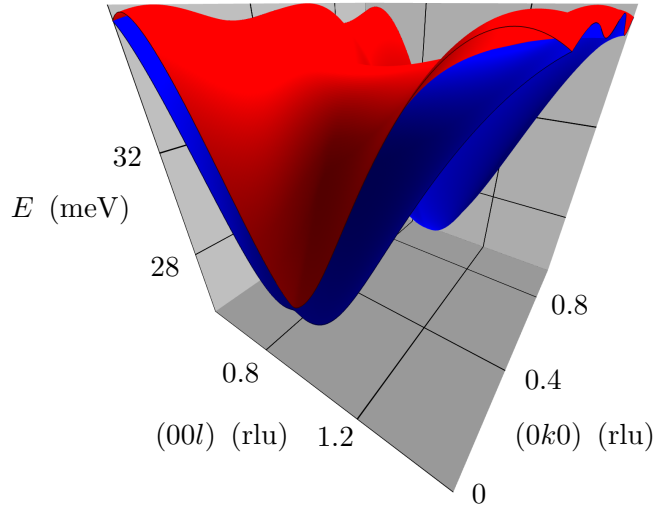
\begin{figure}[htb!]
\begin{centering}
\begin{tikzpicture}
	\draw (0, 0) node [ inner sep = 0 ] { \includegraphics[width=0.45\textwidth]{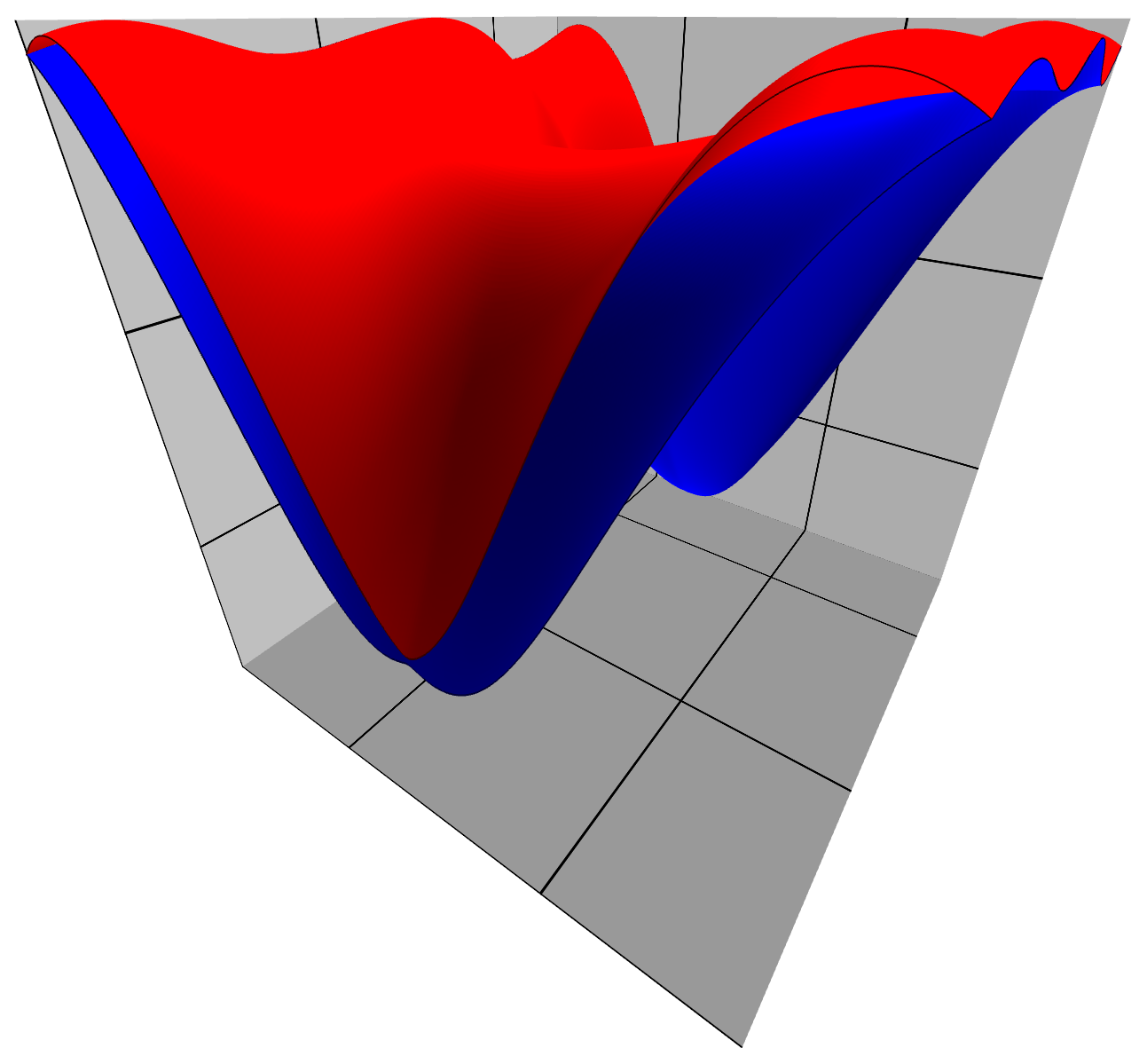} };

	\draw (-4., 0.5) node {  $E\ \left(\mathrm{meV}\right)$ };
	\draw (-2.75, -0.25) node {  $28$ };
	\draw (-3.25, 1.2) node {  $32$ };

	\draw (-2., -2.5) node {  $(00l)\ \left(\mathrm{rlu}\right)$ };
	\draw (-1.75, -1.65) node {  $0.8$ };
	\draw (-0.5, -2.5) node {  $1.2$ };

	\draw (2.8, -2.5) node {  $(0k0)\ \left(\mathrm{rlu}\right)$ };
	\draw (1.4, -3.25) node {  $0$ };
	\draw (2.25, -1.75) node {  $0.4$ };
	\draw (2.6, -0.75) node {  $0.8$ };
\end{tikzpicture}
	\caption{Example dispersion relation calculation for the magnetic configuration of MnTe given in Ref. \cite{Liu2024}.
	In the 3-d plot, the splitting of the two magnon modes and its restriction to specific momenta is well observed.}
	\label{fig:splitting}
\end{centering}
\end{figure}

Having the magnetic structure set up, the dispersion relations can be calculated in 2-d or 3-d.
The altermagnetic spin-splitting is well observed along the $\left(\frac{1}{3},\ 0,\ l\right)$ magnon branch.
Fig. \ref{fig:splitting} shows an example plot produced for this configuration. 
Further plots for the same system are depicted in Fig. \ref{fig:screenshot}.

\section{Impact}
The main advantage of \textit{Magpie} compared to other linear spin-wave calculators
is the visual nature of its design and the efficiency of optimised natively compiled code.
Magnetic models can be created and changed graphically without the need to write scripts.
The effects on the dispersion are visualised immediately and can be inspected interactively.
As such, the software has proven to be useful in teaching new students linear spin-wave theory.

\textit{Magpie} has already been used in several papers \cite{CSO, mnte, Lenander2026, Mecoli2026}, a BSc. thesis \cite{Hertz2026}, 
and several ongoing research projects.
The software is furthermore included in the \textit{Takin} software package where
it replaces its now removed minimalist predecessor \cite{Takin2023} and continues to serve
as magnon calculation core for resolution convolution simulations and fits.
Each of these use cases emphasise different aspects of the software, some of which are described in more detail below.

As first scientific case, \textit{Magpie} was employed to analyse the non-reciprocal magnons
in the field-polarised phase of $\mathrm{Cu}_2\mathrm{OSeO}_3$ \cite{CSO}.
``Non-reciprocal'' means that the dispersion is asymmetric and magnon creation happens at different energies than annihilation.
In $\mathrm{Cu}_2\mathrm{OSeO}_3$, the shape of the dispersion can be switched using an external magnetic field.
The software was used to visualise this behaviour and topological properties like the field-dependence of a band crossing point in 3-d.

As second use case, the intensities of the magnon spectra at the altermagnetic splitting in $\alpha$-MnTe were calculated \cite{mnte}.
This system is also the basis of most of the examples in the present paper and was introduced in Sec. \ref{sec:example}.
For MnTe, a field-dependent change of the magnon intensity was measured using inelastic neutron scattering.
The software was used to calculate and visualise the rotation of the compound's N\'eel vector,
which was found to be the cause of the observed change of magnon intensity.

Further use cases included the analysis of the spin dynamics of $\mathrm{Bi}_2\mathrm{Fe}_4\mathrm{O}_9$\cite{Lenander2026},
as well as a calculating a magnon spectrum for subsequent convolution with the instrumental
resolution function of a time-of-flight spectrometer \cite{Mecoli2026}.
Both these examples show the interoperability of the software.
In the first example, the software was used as part of the neutron scattering analysis software \textit{Takin} \cite{Takin2023}
and in conjunction with the spin-wave calculator \textit{SpinW} \cite{Toth2015} via its ability of import and export models.
In the latter example, \textit{Magpie} was used to help develop a new resolution calculation method,
also in conjunction with \textit{Takin} \cite{Takin2023}.

\subsection{Benchmarks}
\begin{figure}[htb]
\begin{centering}
	\begin{tikzpicture}
		\draw (0, 0) node [ inner sep = 0 ] { \includegraphics[width=0.32\textwidth]{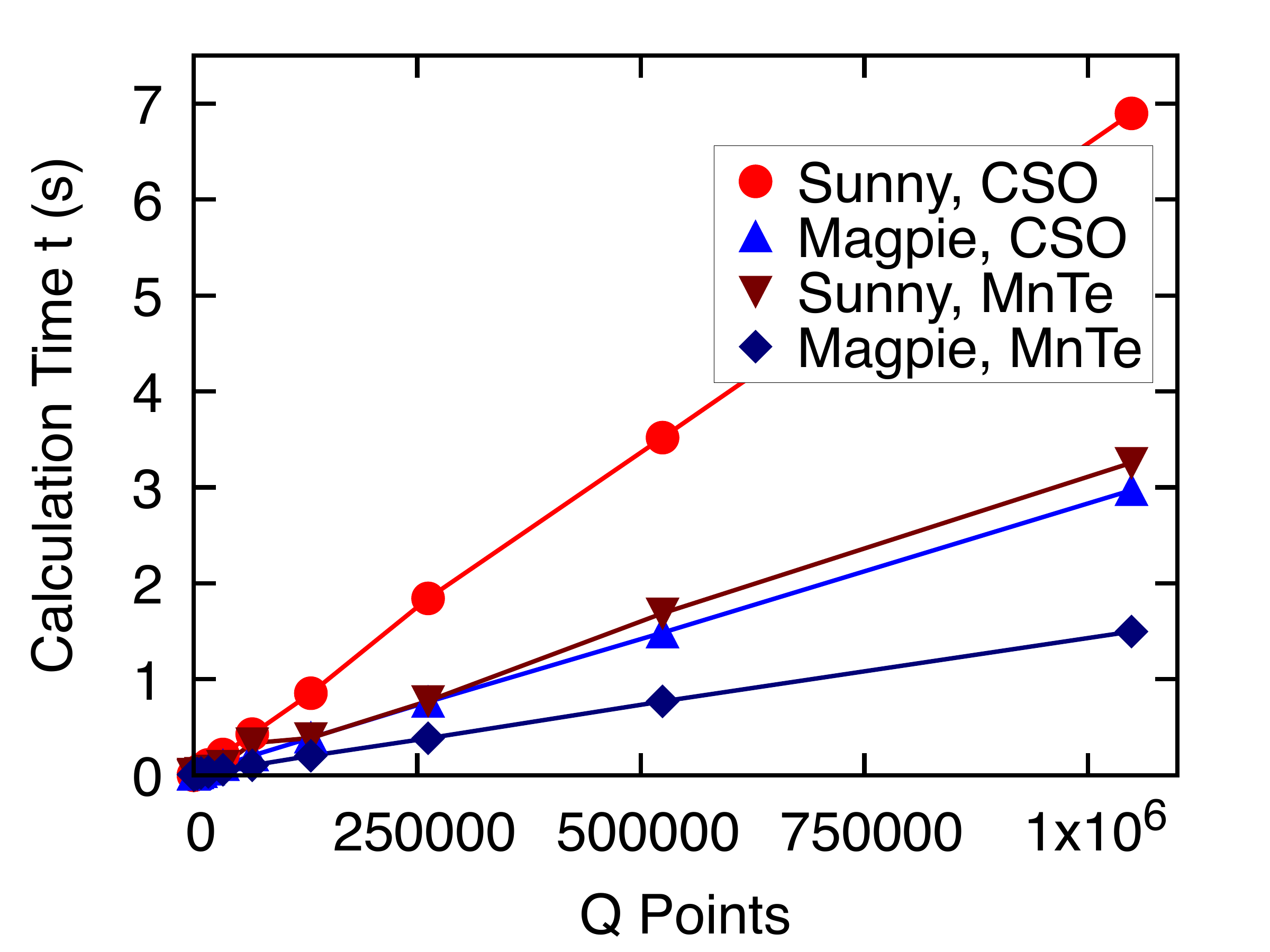} };
		\draw (-0.65, 0.8) node [ inner sep = 0 ] { \includegraphics[width=0.15\textwidth]{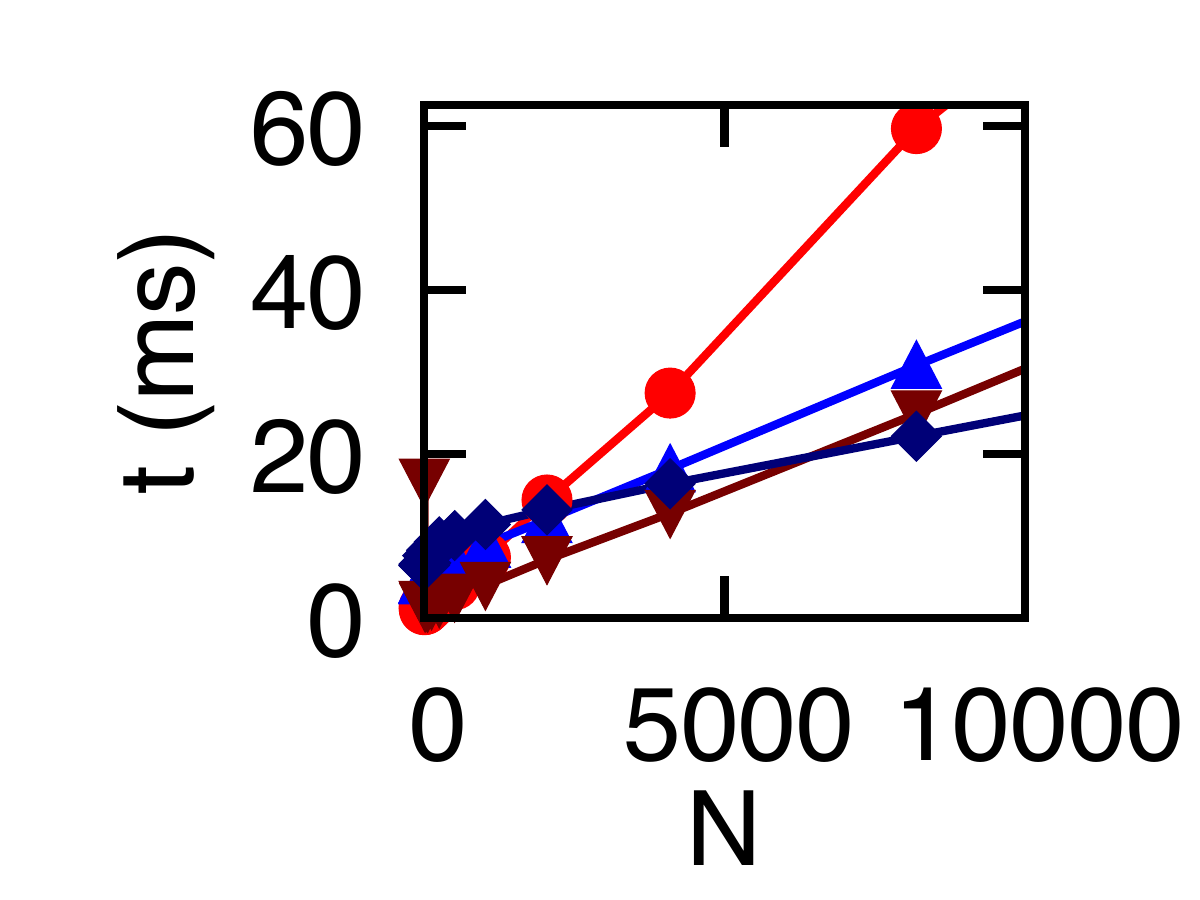} };
		\draw (-2.25, 1.5) node { \bf \fontfamily{phv} \selectfont (a) };
	\end{tikzpicture}
	\begin{tikzpicture}
		\draw (0, 0) node [ inner sep = 0 ] { \includegraphics[width=0.32\textwidth]{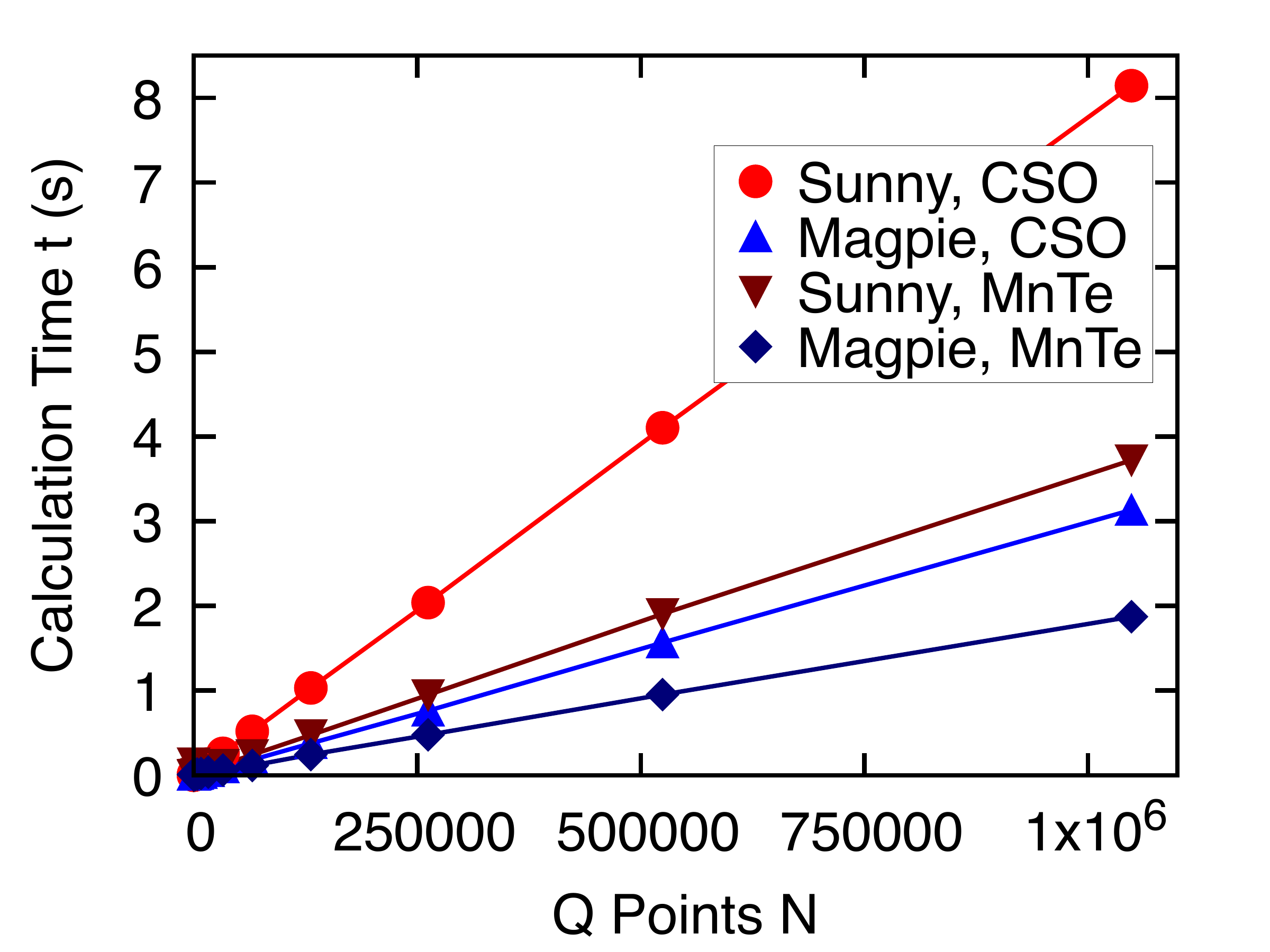} };
		\draw (-0.65, 0.8) node [ inner sep = 0 ] { \includegraphics[width=0.15\textwidth]{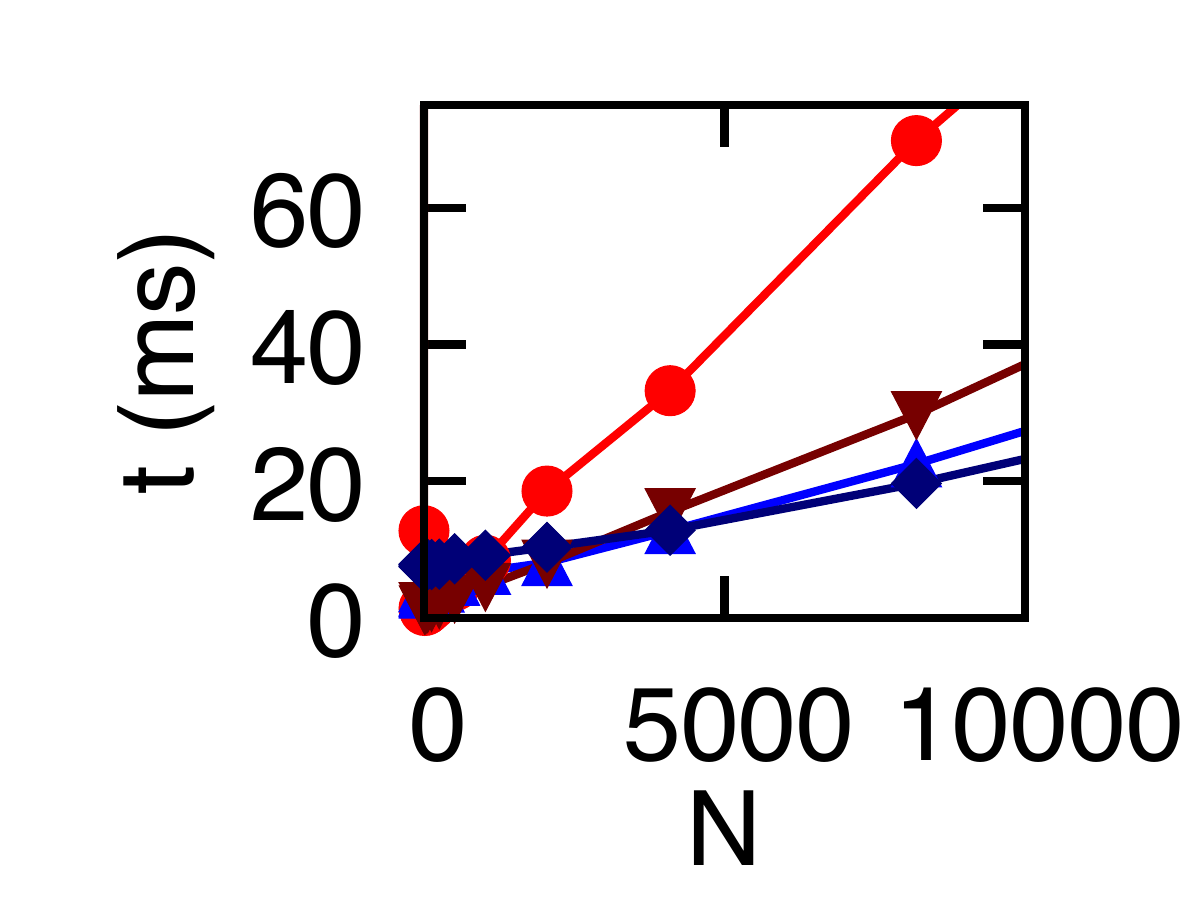} };
		\draw (-2.25, 1.5) node { \bf \fontfamily{phv} \selectfont (b) };
	\end{tikzpicture}
	\begin{tikzpicture}
		\draw (0, 0) node [ inner sep = 0 ] { \includegraphics[width=0.32\textwidth]{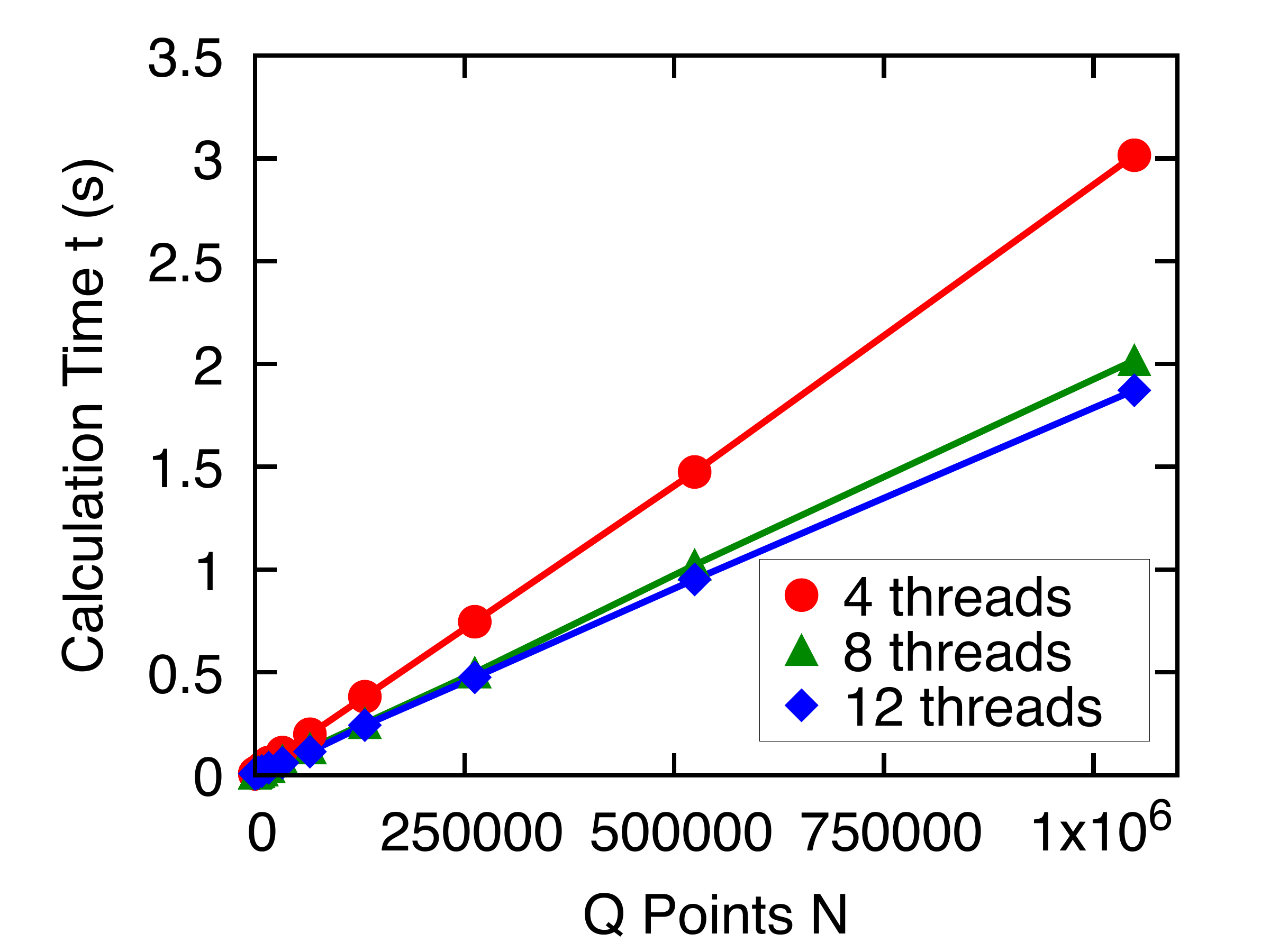} };
		\draw (-0.4, 0.8) node [ inner sep = 0 ] { \includegraphics[width=0.15\textwidth]{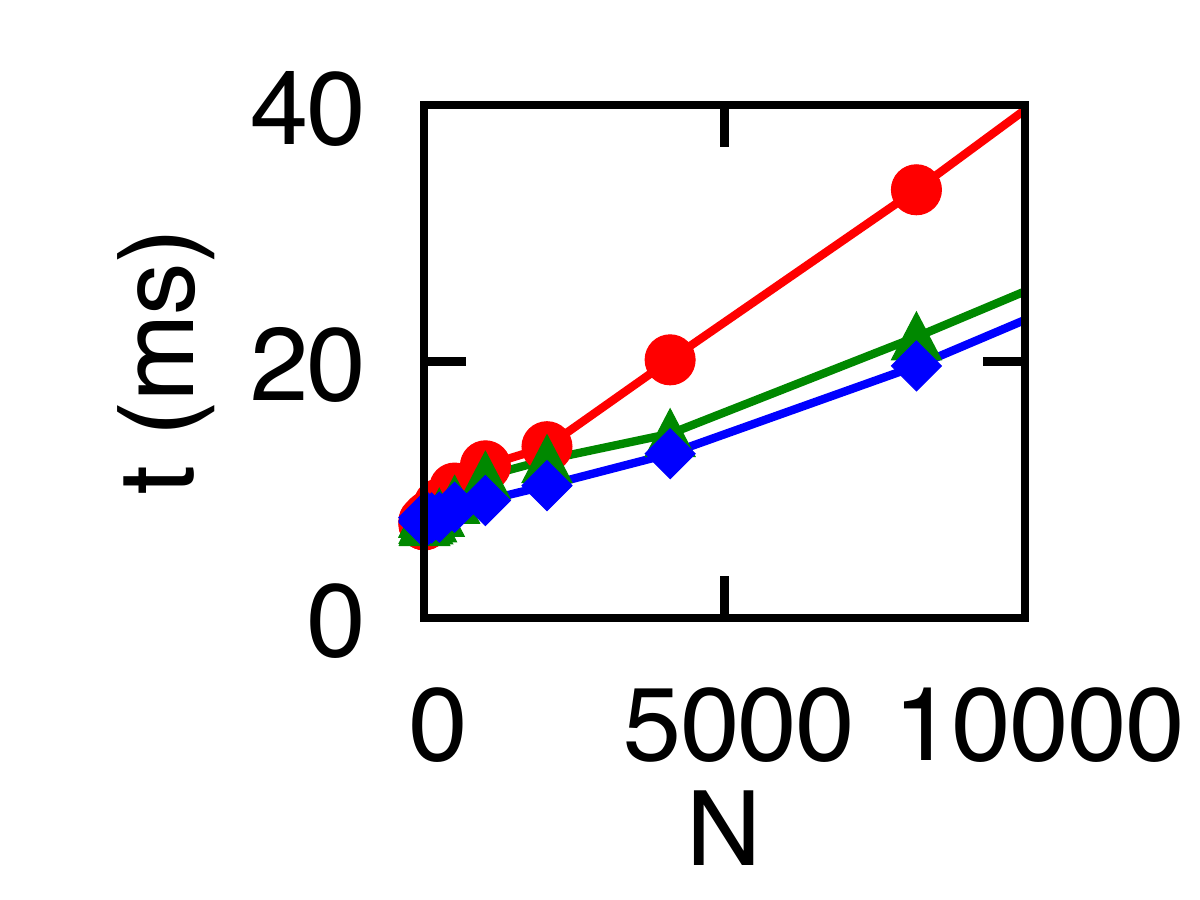} };
		\draw (-2.25, 1.5) node { \bf \fontfamily{phv} \selectfont (c) };
	\end{tikzpicture}
	\caption{(a) Benchmarks performed on an \textit{M4-Pro} 14-core CPU from 2024 using 12 threads.
	(b) Benchmarks performed on an \textit{Ultra 7 265} 20-core CPU from 2025 using 12 threads.
	Both benchmarks, (a) and (b), compare the performance of \textit{Magpie} version 1.2 to \textit{Sunny} \cite{Sunny} version 0.9.2
	using two different magnetic models, namely $\mathrm{Cu}_2\mathrm{OSeO}_3$ (``CSO'') \cite{Luo2020, CSO} and MnTe \cite{Liu2024, mnte}.
	(c) \textit{Magpie} benchmarks performed on the \textit{Ultra 7 265} CPU using the MnTe model and 4-12 threads.
	The insets in the three panels show the respective benchmarks at a lower number of $Q$ points.}
	\label{fig:benchmarks}
\end{centering}
\end{figure}
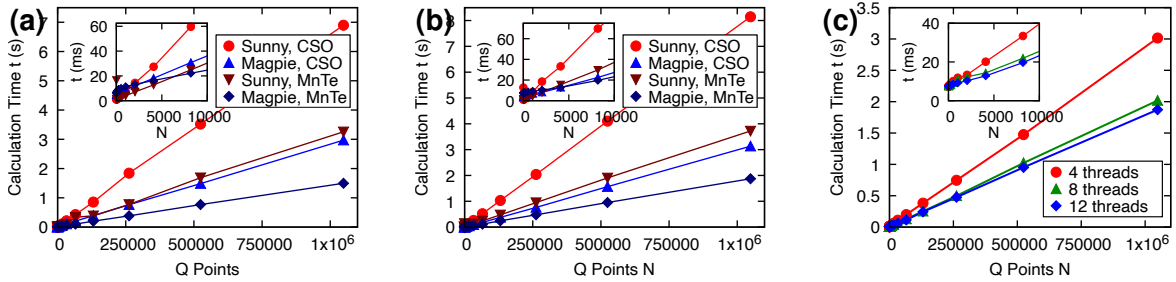

Using the magnetic models from the recent papers mentioned above,
namely altermagnetic MnTe \cite{Liu2024, mnte} and non-reciprocal $\mathrm{Cu}_2\mathrm{OSeO}_3$ \cite{Luo2020, CSO},
benchmarks were performed, measuring the calculation time versus the number of given momentum transfers $Q$.
The results are shown in Fig. \ref{fig:benchmarks}, where they are also compared to the times needed
for the same calculations using the \textit{Sunny} software \cite{Sunny}.
The comparison models used for \cite{Sunny} were generated from the \textit{Magpie} models using its code generator.

The benchmarks were done on two systems, namely on a 14-core \textit{M4-Pro} CPU,
as well as on a \textit{Ultra 7 265} 20-core CPU.
On the two systems, the software was built using \textit{Clang} \cite{clang} and \textit{GCC} \cite{GCC}, respectively.
For the \textit{Magpie} tests, we used 12 threads,
while for \textit{Sunny} -- unlike \textit{Magpie} -- the number of calculation threads cannot be 
freely chosen\footnote{At least not at the moment of writing. An effort to have explicit control over the execution threads seems to be under way in \textit{Sunny}, too.}.
Instead the software relies on the automatic threading of its \textit{Lapack} and \textit{BLAS} backends,
which we manually hinted to use 12 threads using the command \textit{BLAS.set\_num\_threads(12)}.

The respective results for both CPUs are shown in Fig. \ref{fig:benchmarks}, panels (a) and (b).
On both CPUs, \textit{Magpie} is significantly faster than \textit{Sunny}, especially for a high number of $Q$ points.
At a low number of $Q$ points (insets in Fig. \ref{fig:benchmarks} (a), (b)),
\textit{Sunny} is slightly faster, but here we also neglect its significant initial startup time
that the \textit{Julia} interpreter needs for just-in-time compilation.

Fig. \ref{fig:benchmarks} (c) shows \textit{Magpie} benchmarks for the MnTe model using 4, 8, and 12 threads on the \textit{Ultra 7 265} CPU.
Running the calculation in 8 threads yields about 50\% performance gain compared to 4 threads.
Using 12 threads, however, is only slightly faster than 8 threads for this CPU, namely about 8\%.
The reasons are the increasing overhead due to thread and data management and synchronisation.
Given that \textit{Sunny} runs slightly slower than \textit{Magpie} limited to four threads, 
it might be that \textit{Sunny} or its linear algebra library already hits this kind of performance 
limit due to overhead at four threads on the tested CPUs.

\section{Conclusions and Outlook}
We presented a highly performant, modular, and easy to use magnon calculation and visualisation software 
with a fully featured graphical user interface and interconnectivity to other popular spin-wave libraries.
Aside from its use in the research of magnetic excitations, we hope that graphical nature of the software proves useful
in education, helping in introducing new students to the theory of spin-waves.

A currently ongoing project \cite{Hertz2026} is the inclusion of \textit{Magpie} as a backend module for magnon calculations 
in the neutron ray-tracing software \textit{McStas} \cite{McStas}.
This will provide a direct possibility to simulate accurate convolutions of magnetic excitations with instrumental resolutions.
A future development will implement a generalised version of the Landau-Lifshitz formalism \cite{Garst2017}
that we have already used in special cases like helical \cite{Weber2018a, Weber2019Coni} and multiple-propagation
vector systems like magnetic skyrmions \cite{Weber2022Skyrmi, Weber2026}.
In connection with these topological magnons, we will furthermore continue developing the topology module
that is currently still in its testing phase.

\section*{Acknowledgements}
Thanks to Arthur T. Hertz for pointing out the different possible pre-factors in the definitions of single-ion anisotropy.
We thank Marie-Bernadette Lepetit, Elisa Rebolini, Henrik Jacobsen, Victor Mecoli,
Martin Rotter, Duc Le, and Niclas Heinsdorf for many useful discussions.
Thanks to Yash Tijare and Sawyer Beltz for testing the software and to Gabriel Rebolini and Hugo Ponsin for internship work
testing the \textit{Python} scripting interface and embedding \textit{Magpie} in other software projects.

No artificial intelligence has been used in the development of the software.



\section*{Current executable software version}
Executables for various systems can be downloaded under:
\url{https://www.ill.eu/en/for-ill-users/software-and-scientific-tools/software-scientific-tools/magpie/}.

The source code is available via DOI
\href{https://doi.org/10.5281/zenodo.16180814}{10.5281/zenodo.16180814} \cite{Magpie}.

\end{document}